\documentclass[a4paper,11pt]{article}
\pdfoutput=1
\usepackage{jheppub}
\usepackage{bbold}
\usepackage[T1]{fontenc}

\usepackage{float}
\usepackage{comment}

\def\aeff{a_{e}}
\def\reff{r_{e}}
\def\beff{b_{e}}
\def\geff{g_{e}}
\def\Feff{\mathcal{F}_{e}}
\def\Seff{S_{eff}}
\def\cneff{c_{n,e}}
\def\ckeff{c_{k,e}}

\def\half{\frac{1}{2}}

\def\F{\mathcal{F}}

\def\lr#1{\left(#1\right)}
\def\slr#1{\left[#1\right]}

\def\trl#1{\textrm{Tr}\lr{#1}}

\def\CPn{\mathbb C P^n}

\def\RS2{\mathbb R\times S^2_F}

\def \be  {\begin{equation}}
\def \ee  {\end{equation}}
\def \bex  {\begin{equation*}}
\def \eex  {\end{equation*}}
\def \bea {\begin{eqnarray}}
\def \eea {\end{eqnarray}}
\def \bal {\begin{align}}
\def \eal {\end{align}}

\def\pd#1#2{\frac{\partial #1}{\partial #2}}

\def \PRD {{Phys. Rev. D\ }}
\def \JHEP {{J. High Energ. Phys.\ }}

\allowdisplaybreaks
\title{\boldmath Beyond second-moment approximation in fuzzy-field-theory-like matrix models}

\author[a]{M\'aria \v Subjakov\'a}
\author[a]{Juraj Tekel}

\affiliation[a]{Department of Theoretical Physics, Faculty of Mathematics, Physics and Informatics,\\ Comenius University, Mlynsk\'a Dolina, Bratislava, 842 48, Slovakia
}

\emailAdd{maria.subjakova@fmph.uniba.sk}
\emailAdd{juraj.tekel@fmph.uniba.sk}

\abstract{
We investigate the phase structure of a special class of multi-trace hermitian matrix models, which are candidates for the description of scalar field theory on fuzzy spaces. We include up to the fourth moment of the eigenvalue distribution into the multi-trace part of the probability distribution, which stems from the kinetic term of the field theory action. We show that by considering different multi-trace behavior in the large moment and in the small moment regimes of the model, it is possible to obtain a matrix model, which describes the numerically observed phase structure of fuzzy field theories. Including the existence of uniform order phase, triple point, and an approximately straight transition line between the uniform and non-uniform order phases.
}

\begin{document} 
\maketitle
\flushbottom

\section{Introduction}\label{sec_intro}

One of the universal consequences of the future successful combination of quantum theory and theory of gravity is believed to be a fundamental change in the notion of space-time \cite{doplicher}. One particular way to realize this idea is to introduce  noncommutative spaces \cite{cones}. Their appeal is in the fact that they retain at least some of the continuous symmetries of the standard, commutative, space and the fundamental distance provides a natural UV regulator \cite{sF22}. They appear also as effective description of phenomena in string theory \cite{ppsug,witten} and condensed matter physics \cite{qhe2} and compact versions of noncommutative spaces, the fuzzy spaces, arise as solutions in various matrix formulations of string theories \cite{matrixmodels1,steinacker_review}.

The hallmark of the field theories on noncommutative spaces is the existence of a new phase in the phase diagram, the striped or non-uniform order phase. In this phase, the field does not oscillate around a single value in the whole space but rather forms stripes of oscillations around different minima of the potential. This phase exists together with the two standard, commutative phases - the disorder phase, in which the field oscillates around the zero value of the potential, and the uniform order phase, where the field oscillates around one of the minima of the potential. The existence of the two commutative phases has been established in \cite{comr2} and the transition line has been obtained numerically, most recently in \cite{com_num}. The existence of the new phase in noncommutative theories has been shown in \cite{NCphase1} and has been since observed numerically in many different works for the fuzzy sphere \cite{num09,num14,samo}, and other spaces \cite{num_disc,num_RSF2,num14panero2}.

The non-uniform order phase spontaneously breaks the translation symmetry of the space. This is possible even in the two-dimensional case, due to the fundamental nonlocality of the theory. Moreover, the phase does survive the commutative limit of the theory and is thus related to the phenomenon of UV/IR mixing \cite{uvir1,uvir2}.

It has been understood for quite some time now that the best analytical tool to study the properties of field theories on noncommutative spaces are matrix models \cite{steinacker05,PNT12}. Especially for fuzzy spaces, where the finite volume of the space ensures that the theory has only a finite number of degrees of freedom. Real scalar fields on fuzzy spaces are hermitian matrices and the field theory defined through functional integral correlation functions is a specific random matrix model. Calculations in this matrix model are however more complicated than the standard cases since the kinetic term of the field theory depends on the angular variables of the matrix. The numerical analysis of these models led to most of the results for the phase structure of the fuzzy field theories. Moreover, recent simulations of the matrix model describing a version of noncommutative field theory that is free of the UV/IR mixing showed signs of retreat of the non-uniform order phase \cite{belgrade1,belgrade2}, further confirming the connection between the existence of this phase and the UV/IR mixing.

Our main goal in this paper is to reproduce the results of the most recent numerical investigation of the theory on the fuzzy sphere \cite{samo}. In \cite{MSJT2020} we have presented an analysis of a multi-trace matrix model which includes a particular function of the second moment of the eigenvalue distribution and approximates the theory \cite{poly13}. It was shown that certain qualitative features of the phase diagram in the vicinity of the origin of the parameter space are recovered by the model: most importantly the existence of the three phases, transition lines between them, and the existence of the triple point where the three transition lines meet. It was however shown that the phase transition line between the two ordered phases behaves differently in this model for larger values of the parameters. Also, the location of the triple point did not match the predicted location exactly.

To improve on these shortcomings of the model, we include higher moments into the multi-trace part of the probability distribution. In Section \ref{sec2} we give a short overview of the fuzzy field theories and their description in terms of various matrix models. In Section \ref{sec3} we give the necessary details about the calculation of the multi-trace part of the probability distribution in the matrix model stemming from the kinetic term of the field theory. In Section \ref{sec4} we give the outline of our main approach in this paper, a perturbative large parameter solution of the matrix model, and completion of these results using Pade approximants, with more technical details postponed to the appendix \ref{app_direct}. In Section \ref{sec5} we show how the approach works for a naive matrix model, which takes the kinetic as a perturbation to the potential part of the action, and what are the flaws of this model. Finally, in Section \ref{sec6} we show that when one considers a different behavior of the multi-trace part of the action in small and large moments regime, some of the above goals can be met. We add, by hand, a term proportional to the logarithm of the fourth moment, which leads to a model with a better behaved transition line between the symmetric two-cut and asymmetric one-cut phases and preserves the existence of a triple point. Finally, we discuss some of the more complicated modifications one can make in the large moment behavior of the kinetic term effective action.

\section{Matrix models of fuzzy field theories}\label{sec2}

Noncommutative spaces can be defined by the commutation relations of their coordinate functions
\begin{align}
[x_i,x_j] = i\theta^{ij} \ ,
\end{align}
with anti-symmetric ${\theta^{ij}}$ uniquely specifying the space.

In this paper, we will consider only the case of fuzzy sphere \cite{sF21}, which is one of  the simplest examples of the noncommutative geometry. The fuzzy sphere is a compact noncommutative space with the following commutation relation among its coordinates
\begin{align}
[x_i,x_j] = i\theta \epsilon_{ijk} x_k \label{commut_rel} \ .
\end{align}
The algebra of functions on the fuzzy sphere can therefore be identified with the  matrix algebra spanned by three generators $L_i$ of $su(2)$ in the $N$ dimensional representation. By defining
\begin{align}
x_i = \frac{2R}{\sqrt{N^2-1}} L_i \ ,
\end{align}
 where $N$ is the dimension of the matrices and constant $R$ is equal to the radius of the sphere 
  \begin{align*}
 \sum_{i=1}^{3} x_ix_i = R^2\ ,
 \end{align*}
 we satisfy the commutation relations (\ref{commut_rel}), with \begin{align}
 \theta = \frac{2R}{\sqrt{N^2-1}}\ .
 \end{align}
 Taking ${N \rightarrow \infty}$ thus recovers the ''ordinary'' commutative sphere. We will take the sphere of a unit radius ${R=1}$ from now on.
 
 The euclidean scalar field theory on the fuzzy sphere can be stated in terms of the correlation functions
 \begin{equation}
 \langle O(M) \rangle = \frac{1}{Z}\int dM e^{-N^2 S[M]} O[M] \label{correlation}
 \end{equation}
with the integration over all the possible configurations of the hermitian matrix $M$. Therefore, it is equivalent to a random matrix model with the probability measure given by the action of the theory.

The theory of our interest is given by the action \cite{bal}
\begin{equation}
S[M] = \frac{1}{N}\text{Tr} \bigg(\frac{1}{2}r M^2+ g M^4 + \frac{1}{2} M \mathcal{K} M \bigg)
\end{equation}
with the kinetic term equal to the quadratic Casimir invariant of $su(2)$
\begin{equation}
\mathcal{K}M= [L_i,[L_i,M]]\ .
\end{equation}  
This model has been extensively studied by the numerical simulations, showing very different properties than the field theory on the commutative sphere, even as one takes the commutative limit of the underlying space.  

\begin{figure}
    \centering
    \includegraphics[scale=0.4]{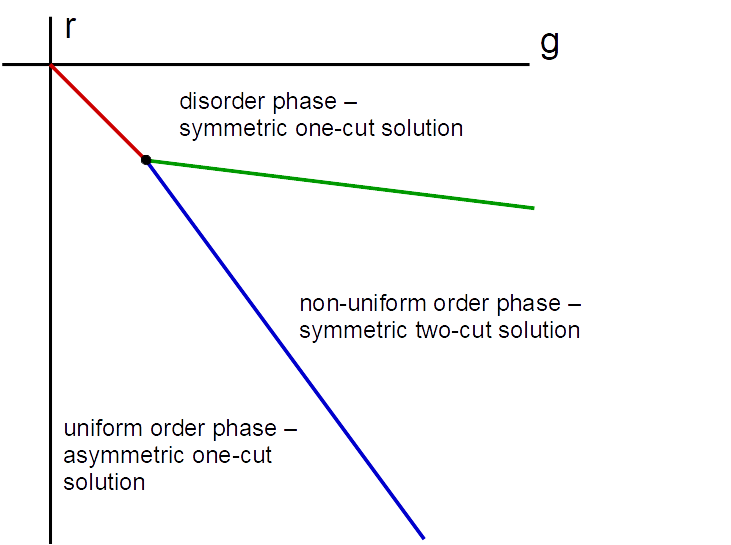}
    \caption{A generic diagram of a noncommutative scalar field theory obtained in various numerical studies. The three transition lines meet at the black triple point. Commutative field theories lack the non-uniform order phase and the green transition line, standard matrix models without the kinetic term lack the uniform order phase and the blue transition line.}
    \label{fig:diagram}
\end{figure}

As we described in the introduction, the scalar field theory on the commutative sphere has two types of solutions. A symmetric one, when the field oscillates around zero, and an asymmetric one, when the field oscillates around one of the minima of the quartic potential. Which solution is realized depends on the values of the parameters $r$ and $g$ in the action.

On the other hand, it was shown that the scalar field theory on the fuzzy sphere has yet another solution besides the two aforementioned cases. In this solution, the field does not oscillate around one value on the whole sphere. Instead, it oscillates around the different potential minima in the distinct sections of the space. 

This is a direct consequence of the so-called UV/IR mixing. As the coordinates on the fuzzy space do not commute, the uncertainty principle does not allow us to measure the position with arbitrary precision. The more precisely we determine one coordinate, the less certain we are in estimating the others. The small scales, therefore, blend with the large ones, hence UV/IR mixing. 

To treat the integrals (\ref{correlation}) analytically, the standard procedure \cite{matrixmodels} is to diagonalize the matrix $M$
\begin{align}
M = U\Lambda U^{\dagger}\ , \qquad \Lambda = \text{diag} (\lambda_1, \lambda_2, \ldots, \lambda_N)\ , \qquad U \in U(N)
\end{align}  
and rewrite the integral in these new variables. For a function ${O[M]}$ which depends only on the eigenvalues $\lambda_i$, we get\footnote{We will assume that the eigenvalues $\lambda$ and the parameters $r$ and $g$ are scaled in such a way that all the terms in the probability distribution contribute and are of the order $N^2$. This is dictated by the logarithmic Vandermonde term, which does not change with any scaling. Since there are three other terms and three parameters to scale, such scaling is possible.}
\begin{align}
\langle O[\Lambda ]\rangle & = \frac{1}{Z}\int \bigg( \prod_{i=1}^N \lambda_i \bigg) O[\Lambda] e^{-N^2\big[ 
\frac{1}{2}r \frac{1}{N}\sum_i \lambda_i^2 + g\frac{1}{N} \sum_i \lambda_i^4 - \frac{2}{N^2}\sum_{i < j} \log|\lambda_i-\lambda_j| \big] } \times \nonumber \\
&\times \int dU e^{- N \text{Tr}\big[\frac{1}{2}U\Lambda U^{\dagger}\mathcal{K}(U\Lambda U^{\dagger})\big]}\ .
\end{align}
The logarithmic term in the action comes from the Jacobian determinant due to the change of the variables. Notice that the kinetic term is not invariant under the unitary transformations, which leads to a non-trivial angular integration
\begin{align}
\int dU e^{- N \text{Tr}\big[\frac{1}{2}U\Lambda U^{\dagger}\mathcal{K}(U\Lambda U^{\dagger})\big]} \label{angular}
\end{align} 
that has not yet been performed analytically. Thus, the integral (\ref{angular}) presents the key difficulty in obtaining the analytical solution of the model, as we are unable to get the full analytical solution and are left only with the possibility of approximate results.

\section{Angular integral as a multi-trace matrix model}\label{sec3}

In this section, we review the approximations of the kinetic term  (\ref{angular}) that have been studied in recent years, as well as the solutions of such approximative models. 
 
Let us denote
\begin{equation}
e^{-N^2S_{eff}} = \int dU e^{- N \text{Tr}\big[\frac{1}{2}U\Lambda U^{\dagger}\mathcal{K}(U\Lambda U^{\dagger})\big]}\ ,\label{sec3:effaction}
\end{equation}
we will, therefore, look into the possible approximate formulas for the effective action $S_{eff}$.

\subsection{Perturbative expansion}\label{sec3.1}

The most straightforward approximation is to consider a perturbative expansion of the angular integral (\ref{angular}) in a small $\epsilon$ parameter \cite{ocon,samann,samann2,samanfuzzydisc}
\begin{align}
 \int dU e^{- \epsilon \text{Tr}\big[ \frac{1}{2}U\Lambda U^{\dagger}\mathcal{K}(U\Lambda U^{\dagger})\big]} \,=&\, 1 - \frac{\epsilon}{2} \int dU  \text{Tr}\big[U\Lambda U^{\dagger}\mathcal{K}(U\Lambda U^{\dagger})\big]+ \nonumber \\
 &+ \frac{\epsilon^2}{8} \int dU \bigg( \text{Tr}\big[U\Lambda U^{\dagger}\mathcal{K}(U\Lambda U^{\dagger})\big] \bigg)^2 + \ldots\ .
\end{align}
The integrals in the perturbative series are still technically demanding, and the computational difficulty significantly increases with the series order. Nevertheless, the first four orders were obtained \cite{samann2}, leading to the following expression for the effective action
\begin{align}\label{3_pert_sphere}
S_{eff} = \frac{1}{2} \bigg( \frac{1}{2} t_2- \frac{1}{24}t_2^2+ \frac{1}{2880}t_2^4\bigg) - \frac{1}{432}t_3^2- \frac{1}{3456}(t_4-2t_2^2)^2+ \ldots\ , 
\end{align}
where $t_n$ are symmetrized moments\footnote{We discuss relationship between the standard moments $c_n=\trl{M^n}/N$ and the symmetrized moments $t_n$ in the appendix \ref{app_direct}.}
\begin{align}
t_n = \frac{1}{N} \textrm{Tr}\,\lr{M- \mathbb{1} \frac{1}{N}\textrm{Tr}\,M }^n\ .  
\end{align}
Note that this is a multi-trace expression. The kinetic term in the action, therefore, effectively contributes with the multi-trace terms and, as such, leads to a multi-trace matrix model.  

However, such a perturbative model does not capture the key features of the theory, known from the numerical simulations, as is discussed in Section \ref{sec5}. 

\subsection{Second moment model} \label{sec3.2}

In this section, we describe the approximation that has so far been the most successful in capturing the main properties of the scalar field theory. This approximation is based on the fact that the free theory (i.e., theory with ${g=0}$) is analytically fully solvable, and in such case, the kinetic term in the action effectively rescales the solution \cite{steinacker05,PNT12}.  
 
Thus, one can  divide the effective action $S_{eff}$ into two parts
 \begin{align*}
 S_{eff} = \frac{1}{2} \mathcal{F} + \mathcal{R}\ ,
\end{align*}  
where $\mathcal{F}$ corresponds to the changes the kinetic term causes in the case of the free theory, and $\mathcal{R}$ does not contribute in such case \cite{poly13}. The function $\mathcal{F}$ was determined to be
\begin{align}\label{F2}
\mathcal F(t_2) =  \log \bigg(\frac{t_2}{1-e^{-t_2}} \bigg)
\end{align}
and the remainder term $\mathcal R$ being a function of $t_n$'s, vanishing for a semicircle distribution. Note that this structure is consistent with the perturbative result (\ref{3_pert_sphere}).

The approximation of the kinetic term with only the function $\F(t_2)$ captures some of the key features of the full model, namely the existence and rough location of the triple point. While the value $g_c$ of the critical point obtained by the perturbative solution of this approximation agrees with the value obtained by the most recent numerical simulation, the parameter $r_c$ was greater than the numerical result by approximately a factor of $10$.

The reason why the approximation (\ref{F2}) is so much more successful than the small expansion of the kinetic term (\ref{3_pert_sphere}) discussed in the previous section lies in the behavior of the approximation for the large moments. The polynomial terms in (\ref{3_pert_sphere}) introduce, according to their sign, attraction or repulsion between the matrix eigenvalues that is just too strong to give reasonable results for the small values of $g$, see section \ref{sec5}. The function $\F(t_2)$, however, behaves logarithmically for large moment $t_2$, which seems to generate just the right amount of attraction among the eigenvalues.  

Nevertheless, as we have already mentioned in the introduction, the approximation fails to correctly reproduce all the properties of the full model for larger values of the interaction parameter $g$. Namely, the character of the transition line between the symmetric disorder phase and the asymmetric uniform-ordered phase. In the approximate model, this transition line approaches a finite value of the parameter $g$ as $|r|$ increases to the infinity to the contrary with the numerical simulations, which suggest a linear transition.   

\subsection{Higher moments model}

We shall consider a generalized model, with the kinetic term effective action as follows
\be\label{model}
S_{eff} = F\slr{c_1,t_2,t_3,t_4-2t_2^2}\ , 
\ee
where ${F[y_1,y_2,y_3,y_4]}$ is a well behaved function of four variables. We will denote\footnote{The function $F$ for the matrix models describing the fuzzy field theories will not depend explicitly on the first moment $c_1$ and will only include this variable through combinations of $t_n$'s. But since the approach we describe is more general, we leave this possibility even though in all the cases discussed in this paper $f_1=0$.}
\begin{align}
f_1=\pd{F}{y_1}\ ,\ f_2=2\pd{F}{y_2}\ ,\ f_3=\pd{F}{y_3}\ ,\ f_4=\pd{F}{y_4}\ .
\end{align}
Note the extra factor of $2$ in the definition of $f_2$, which removes confusing factors of $2$ in the expressions for the effective mass parameter in the rest of the text.

This form is a generalization of the perturbative result (\ref{3_pert_sphere}). The complete effective action is a function of all the symmetrized moments and considering such an expression only up to the fourth moment is our main approximation. It is the next natural step to take after the analysis \cite{MSJT2020} has shown that considering only the second moment does not yield results that are sufficient and need improvement.

We will now investigate the phase structure of the multi-trace matrix model (\ref{model}), focusing on the region of parameter space, where the asymmetric one-cut solution is the preferred one.

\section{Large $r$ solution of fourth moment fuzzy-field-theory-like matrix models}\label{sec4}

In this section, we will present the general approach to the solution of the fourth-order multi-trace model given by the effective action (\ref{model}). We will first set up the basic formulae for the three types of the relevant solutions - symmetric one-cut, symmetric two-cut, and asymmetric one-cut. Then we will outline the general approach to solving the phase transition conditions and finding the phase diagram of the model.

\subsection{Setup}\label{sec4.1}

As with any other multi-trace model, the idea of the analysis of the model (\ref{model}) is to convert it to an effective single-trace model with additional self-consistency conditions on the moments of the distribution, for more details see for example \cite{CORFU19a} and for a different approach see \cite{newYdri}. 

The technical details are postponed to the appendix \ref{app_direct} and here we present only the final conditions on the relevant solutions of the model.

\subsubsection*{Symmetric regime}
The {\bf symmetric one-cut solution}  
supported on the interval ${(-\sqrt{\delta},\sqrt{\delta})}$ is determined by the following two conditions for $\delta$ and $\geff$
\begin{align}\label{exp_sym}
\frac{4}{\delta}-\delta(2g+\geff)+\half \delta^3\geff(\geff-g)\,=&\, r+ f_2\slr{0,\frac{\delta}{4}+ \frac{\delta^3\geff}{16},0,-\frac{\delta^4 \geff (2 + \delta^2 \geff)}{128} }\ ,\\ \label{exp_sym0}
\geff \,=&\, g+f_4\slr{0,\frac{\delta}{4}+ \frac{\delta^3\geff}{16},0,-\frac{\delta^4 \geff (2 + \delta^2 \geff)}{128} }\ .
\end{align}
The free energy of such solution is
\begin{align}\label{free_1cut}
\F\,=\,& \Feff +  
\frac{-32 + 8 \delta^2 (g + \geff) + 3 \delta^4 \geff (g + \geff) + 8 \delta r + 
   2 \delta^3 \geff r}{64}+ \nonumber\\& {\color{white}\Feff} + F\slr{0,\frac{\delta}{4}+ \frac{\delta^3\geff}{16},0,-\frac{\delta^4 \geff (2 + \delta^2 \geff)}{128} }\ ,\\
&\Feff\,=\,\frac{3}{8}-\frac{1}{2}\log\bigg(\frac{\delta}{4}\bigg)+ \frac{4-3\delta^2g_e}{384}\lr{36+3\delta^2g_e} \ .
\end{align}

The {\bf symmetric two-cut solution} supported on ${(-\sqrt{D+\delta},-\sqrt{D-\delta})\cup(\sqrt{D-\delta},\sqrt{D+\delta})}$ is determined by two conditions on $D$ and $\delta$
\begin{align}
\frac{1}{\delta^2}\,=\,&g+f_4\slr{0,D,0,\frac{\delta^2}{4}-D^2}\ , \label{exp_2cut}\\
\frac{4D}{\delta^2}\,=\,&r+8Dg+f_2\slr{0,D,0,\frac{\delta^2}{4}-D^2}\ . \label{exp_2cut0}
\end{align}
The free energy is
\begin{align}\label{free_2cut}
\F\,=\,& \Feff + \frac{1}{4}\lr{2 D r + g \delta^2-1 + 4 D^2 \lr{\frac{1}{\delta^2} + g}}+ F\slr{0,D,0,\frac{\delta^2}{4}-D^2}\  \\
&\Feff\,=\,\frac{3}{8}+\frac{1}{4}\log\lr{\frac{4}{\delta^2}}-\frac{D^2}{\delta^2}  \ .
\end{align}

The {\bf symmetric phase transition} between these two solutions can be further simplified. The transition is given by the condition ${\reff=-4\sqrt{\geff}}$ and the expressions for the two-cut solution (\ref{app_2cut}) lead to
\begin{align} 
\geff\,=\,&g+f_4\slr{0,\frac{1}{\sqrt{\geff}},0,-\frac{3}{4\geff}}\ , \label{sym_trans1} \\
r\,=\,&-\frac{8g-4\geff}{\sqrt{\geff}}-f_2\slr{0,\frac{1}{\sqrt{\geff}},0,-\frac{3}{4\geff}}\ . \label{sym_trans2}
\end{align}
The first equation is to be solved for $\geff$, which is then to be used in the second equation.

\subsubsection*{Asymmetric regime}
The {\bf asymmetric one-cut solution} supported on the interval ${(D-\sqrt{\delta},D+\sqrt{\delta})}$ is determined by the following four conditions
\begin{align}
f_1\,=&\, 
 \frac{3}{16} D \delta^3 g \geff (4 + 3 \delta^2 \geff - 
    18 D^2 \delta^3 \geff^2) -4 c_1^3 g-\nonumber\\&-
 c_1 \lr{3 \delta g + \frac{3}{4} \delta^3 g \geff - \frac{27}{4} D^2 \delta^4 g \geff^2 + r}\ ,\\
f_2\,=&\,\frac{4}{\delta} + 18 D^2 \delta^2 \geff^2 + \half \delta^3 \geff (\geff-g )+ \lr{\frac{9}{4} D^2 \delta^4 \geff^2 - 
 \delta}(2 g + \geff)  
- 12 c_1^2 g - r
 \ ,\\
 f_3\,=&\,D \geff (4 + 3 \delta^2 \geff)- 4 c_1 g \ ,\label{exp_asym}\\
f_4\,=\,&\geff-g\ .\label{exp_asym0}
\end{align}
for the first moment $c_1=\frac{1}{N}\trl{M}$ and for ${\delta,D}$ and $\geff$. Here, one needs to use the explicit form of the moments of the distribution (\ref{app_cneff0}-\ref{app_cneff1}) given in the appendix \ref{app_direct}.

The free energy of such solution is
\begin{align}\label{free_asym}
\F\,=\,& \Feff+
\frac{1}{256 \delta} \Big(864 c_1 D^3 \delta^7 g \geff^3 - 
   48 c_1 D \delta^4 g \geff (4 + 3 \delta^2 \geff) - 
   3 D^4 \delta \geff (256 + 81 \delta^8 g \geff^3) +
\nonumber\\& {\color{white}\Feff} +   2 D^2 \big(256 + 576 \delta^2 \geff + 384 \delta^4 \geff^2 - 
      36 \delta^6 g \geff^2 + 27 \delta^8 g \geff^3 - 
      36 \delta^5 \geff^2 (12 c_1^2 g + r)\big) + 
      \nonumber\\& {\color{white}\Feff} +
   4 \delta \big(-32 + 64 c_1^4 g + 8 \delta^2 (g + \geff) + 
      3 \delta^4 \geff (g + \geff) + 8 \delta r + 2 \delta^3 \geff r + \nonumber\\& {\color{white}\Feff} +
      8 c_1^2 (12 \delta g + 3 \delta^3 g \geff + 4 r)\big)\Big)+F\slr{c_1,t_2,t_3,t_4-2t_2^2}\\
&\Feff\,=\,\frac{3}{4}-\frac{2 D^2}{\delta}+3 D^4 \geff-\frac{9}{2} D^2 \delta \geff-\frac{1}{4}\delta^2 \geff-\frac{3}{2} D^2 \delta^3 \geff^2-\frac{3}{128} \delta^4 \geff^2-\half\log\left(\frac{\delta}{4}\right) 
\end{align}

Let us note that there technically are also asymmetric two-cut solutions. However, since such solutions were not observed in numerical simulations, we expect that they always have higher free energy than the asymmetric one-cut solution. This was the case for the second-moment model \cite{JT18} and we will not consider this possibility any further.

\subsection{General approach to solving the model}

We have collected all the necessary conditions and the expressions for the free energies. Our goal now is to investigate the space of the parameters of the original model, which we remind the reader are $r$ and $g$. We want to find the regions of existence of the three different solutions and see which solution is the preferred one where these regions overlap.

To do so, we will solve the equations as a series in the powers of $-1/r$, i.e. in the limit of large and negative $r$. This approach was successfully employed in the analysis of a less complicated second-moment model in \cite{MSJT2020}. The basic idea of the process is to solve the equations to sufficiently high order and then complete the perturbative series using the Pade approximation method.

Note that this perturbative approach is very different from the perturbative approach of sections \ref{sec3.1} and \ref{sec5}. There the kinetic term of the effective action (\ref{sec3:effaction}) is calculated as a perturbative series in powers of the kinetic term contributions. This perturbative expansion reflects our limited knowledge about the kinetic term effective action (\ref{sec3:effaction}). Here we solve the defining equations for the phases of the model as a series in powers of $-r$, due to our limited ability to solve the equations of the previous section analytically.

The idea of this approach is that the moments of the distribution become either very large or very small in this limit. The potential wells become very deep and very far apart, thus the asymmetric one-cut solution will be very narrow and the symmetrized moments will be very small, while the two parts of the two-cut solution will be very separated and yield large symmetrized moments. The situation is a little more problematic for the symmetric one-cut solution and we will discuss this issue in the section \ref{sec6}.

Note that it was important that the multi-trace selfinteraction was small compared to the force due to the potential. As we will see shortly, if this is not the case, one can not use this approach and one needs to solve the equations numerically. However, as we will also see, matrix models where the selfinteraction is large do not yield results consistent with noncommutative field theories.

\section{Solutions of the perturbative model on the fuzzy sphere}\label{sec5}

We first solve the fuzzy sphere model (\ref{3_pert_sphere})
\be\label{5sphere}
S_{eff}=\frac{1}{4} t_2-\frac{1}{48}t_2^2+\frac{1}{5760} t_2^4-\frac{1}{432}t_3^2-\frac{1}{3456}\lr{t_4-2t_2^2}^2\ ,
\ee
where the kinetic part of the field theory action has been taken as a perturbation up to the eight order in the eigenvalues of the matrix. Solution of this model will illustrate the method we will employ throughout the rest of the paper and also highlight some of the main problems we will need to address.

A simpler version of this multi-trace model, which considered the asymmetric term ${\trl{M}\trl{M^3}}$, has been analyzed both numerically and analytically in \cite{ydriMultitrace,ydriMultitrace2,newYdri}. These works have identified the phase diagram in a general agreement with the expectation given in the figure \ref{fig:diagram}. As mentioned in the introduction, our goal is to improve on results such as these by inclusion of the rest of the multitrace terms, including the fourth moment of the eigenvalue distribution, into the effective action\footnote{In \cite{samanfuzzydisc}, authors analyze a multitrace matrix model which is more complicated than ${\trl{M}\trl{M^3}}$ related to the theory on the fuzzy disc. However even this model does not include multitraces of the fourth moment $\trl{M^4}$.}.

Without much elaboration, let us comment briefly on the form of the above action. The multi-trace terms add interaction among the eigenvalues. Looking directly at the action (\ref{5sphere}) we see that, for example, configurations with a large symmetrized moment $t_3$ have a lower value of the action and thus also lower free energy. This makes them energetically favorable and thus such a term with a negative sign in the action tends to favor configurations with larger $t_3$ - it pushes eigenvalues further from the center of the distribution and acts as a repulsive force. Similarly terms with a positive sign act as an attractive interaction, even though this interaction is not of the standard pairwise form. Obtaining very large moments $t_3$ and $t_4-2 t_2^2$ lowers the free energy under any bound and tends to destabilize the whole model in the process. But the details of this are far from clear since we deal with symmetrized moments $t_n$, rather than actual moments $c_n$, which further come in peculiar combinations.

We will thus, in a feynmanian spirit, proceed to solutions of the model. We first analyze the symmetric phase transition which does not require calculation of any solution. Then we analyze the asymmetric one-cut solution, followed by the symmetric two-cut solution. This is where we stop, for reasons that will become clear shortly.

\subsubsection*{Symmetric phase transition}

For the effective action (\ref{5sphere}), the conditions for the symmetric phase transition (\ref{sym_trans1},\ref{sym_trans2}) can be solved analytically and yield an expression which we present for the sake of completeness and the readers' amusement
\begin{align}
    r_{c,sym}=\frac{\left(\begin{array}{c}
         -11520 g^2+10 \sqrt{576 g^2+1}-5 \sqrt{3} \sqrt{\left(576 g^2+1\right) \left(\sqrt{576 g^2+1}+24 g\right)}+\\
         +2-120 g \left(4 \sqrt{576 g^2+1}+\sqrt{3} \sqrt{\sqrt{576 g^2+1}+24 g}-2\right)
    \end{array}\right)}{10 \sqrt{3} \left(\sqrt{576 g^2+1}+24 g\right)^{3/2}}\ .\label{sec5-full}
\end{align}
The interesting aspects of the above formula are the large $g$ and the small $g$ expansion. The large parameter expansion is
\begin{align}
    r_{c,sym}=-4 \sqrt{g}-\frac{1}{2}+\frac{1}{12}\frac{1}{\sqrt{g}}+\frac{7}{5760}\frac{1}{g^{3/2}}+\ldots\ ,
\end{align}
i.e. a modification of the usual matrix phase transition $r=-4\sqrt{g}$, while the small $g$ behaviour is
\begin{align}
    r_{c,sym}=0.193\, -38.8 g+\ldots\ .\label{smallg}
\end{align}
However, as we will shortly see, the interpretation of this result close to the origin of the parameter space is rather complicated. But first, let us deal with the asymmetric solution.



\subsubsection*{Asymmetric one-cut solution}

The relevant equations (\ref{exp_asym}-\ref{exp_asym0}) for the effective action (\ref{5sphere}) are just a set of polynomial equations. Following the outlined method, we look for the solution as a power series in $-1/r$, and up to the fourth-order we obtain\footnote{The order of calculation is in principle limited just by our patience and computing power. One can go quite further without too much trouble, the explicit formulae are however not very illuminating.}
\begin{align}
D & = \frac{\sqrt{-r}}{2 \sqrt{g}} - 3\sqrt{g} \frac{1}{(-r)^{3/2}}+ \frac{1 + 1296 g}{1152 \sqrt{g}(-r)^{5/2}}  - \frac{1 + 6480 g + 746496 g^2}{13824\sqrt{g}(-r)^{7/2}} +\ldots\ , \label{asymD}\\
\delta & = - \frac{2}{r}- \frac{1}{2r^2} + \frac{1+180g}{6r^3} - \frac{1+432g}{16r^4} + \ldots\ , \\
g_e & = g - \frac{g}{864r^4}+ \ldots\ , \\
c_1  & =  \frac{\sqrt{-r}}{2 \sqrt{g}} - \frac{3\sqrt{g}}{2(-r)^{3/2}}+ \frac{3\sqrt{g}}{8(-r)^{5/2}} - \frac{\sqrt{g}+144g^{3/2}}{8(-r)^{7/2}}+ \ldots
\end{align}
with the free energy given by
\begin{align}
 \mathcal{F}_{as1c} & = - \frac{r^2}{16g} + \bigg[ \frac{3}{4} + \frac{1}{2}\log ( -2r) \bigg] - \frac{1}{8r} - \frac{1 + 120g}{48r^2} - \frac{1+276}{192r^3} -\frac{\frac{225 g^2}{8}+\frac{35 g}{48}+\frac{1}{640}}{r^4}+ \ldots\ . \label{F_as1c}
\end{align}
We observe that since $r$ is negative all these are alternating series. This suggests there is a reasonable all order solution, which can be approximated for example by Pade approximation. The model (\ref{5sphere}) thus has a well-behaved asymmetric one-cut solution for negative values of $r$, as expected from a matrix model aspiring to describe the fuzzy field theory.

This can be traced to the fact that the large $r$ solution of the equations for the asymmetric one-cut solution (\ref{exp_asym}-\ref{exp_asym0}) requires the small moment expansion of the function (\ref{model}), which is precisely what we work with here. Things are however not this nice in the symmetric regime, where one needs large moment behavior of (\ref{model}), for which (\ref{5sphere}) fails.

\subsubsection*{Symmetric two-cut solution}

The situation for the symmetric two-cut solution is more tricky. The conditions (\ref{exp_2cut},\ref{exp_2cut0}) for the action (\ref{5sphere}) do not admit large $-r$ solutions and our approach is not applicable in this case.

\begin{figure}
    \centering
    \includegraphics[width=0.48\textwidth]{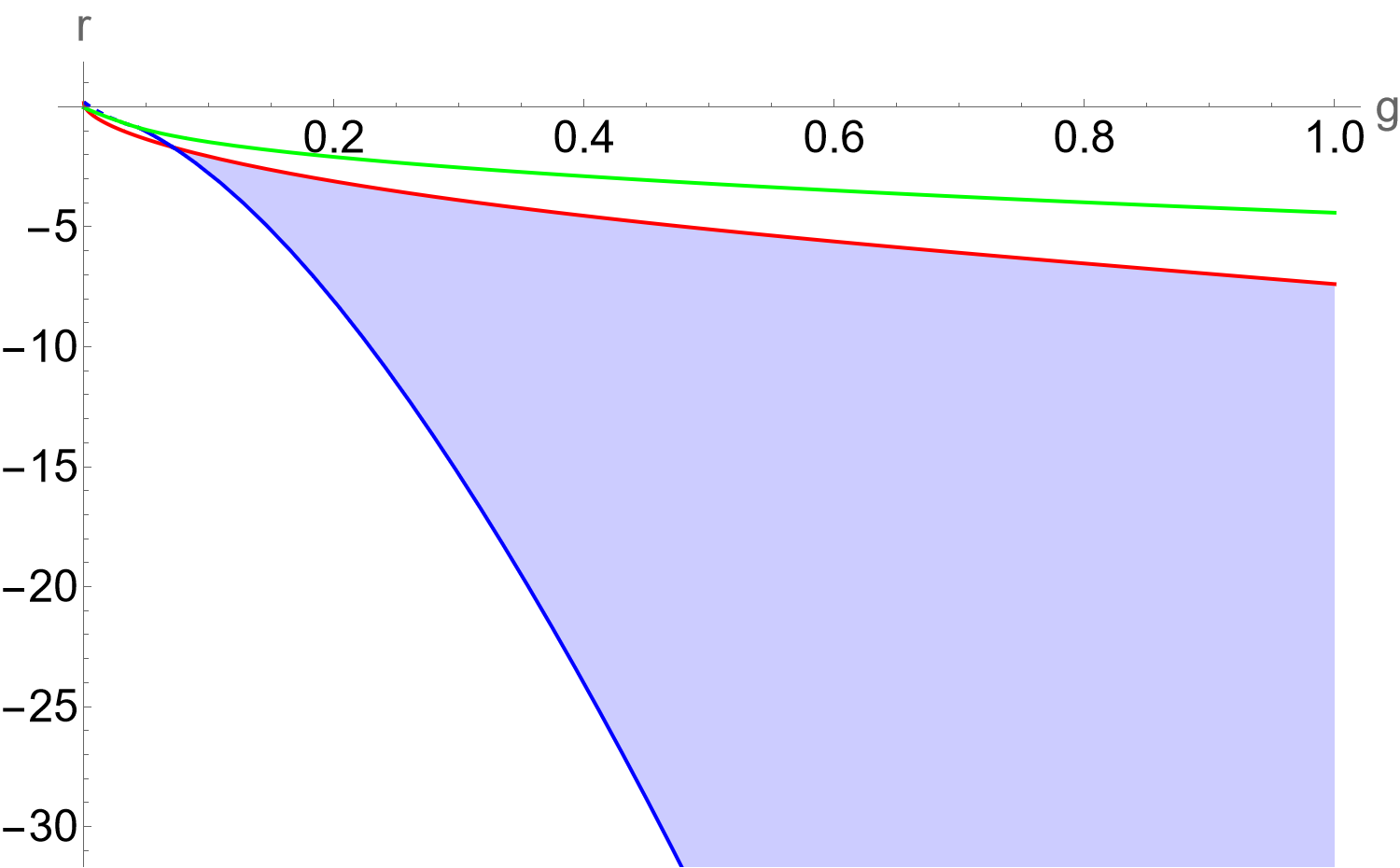}
    \includegraphics[width=0.48\textwidth]{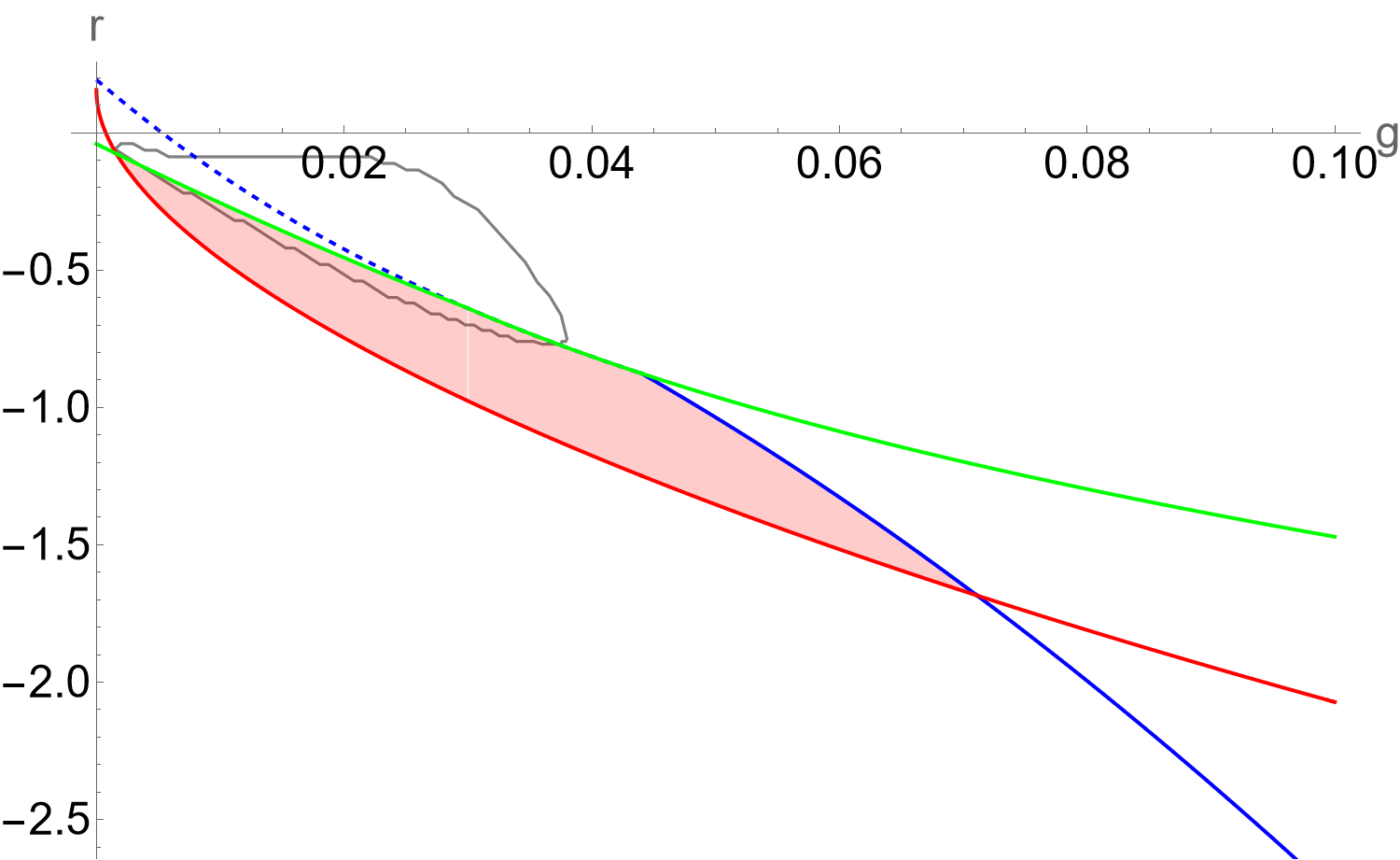}
    \caption{The phase diagram of the multirace matrix model (\ref{5sphere}), with the right image zooming in on the vicinity of the origin of the parameter space. The three lines are boundaries of existence of the three types of solutions. The symmetric one-cut solution exists only above the green line, the symmetric two-cut solution exists only above the blue line and the asymmetric one-cut solution exists only below the red line. For the left image, in the region above the blue and below the red lines, where both two-cut and asymmetric one-cut solutions exist, the free energy of the two-cut solution is always smaller and thus this solution is preferred. See the text for the discussion of the situation in the right image. All lines in these images have been obtained numerically. The dashed blue line and the part of the green line to the right of the intersection with blue line agrees with (\ref{sec5-full}).}
    \label{fig-naive-diagram}
\end{figure}

To study the two-cut solution, we need to investigate the equations (\ref{exp_2cut},\ref{exp_2cut0}) numerically. We scan the parameter space and for given numerical values of $r$ and $g$ try to find a solution to the set of equations defining each of the three phases. If we find a solution, we compute its free energy and move to a different set of values. At the end of the process, we compare the free energies in regions, where more than one solution exists. There are even values of parameters where two solutions of the same kind, e.g. two two-cut solutions, compete. The resulting phase diagram is given in the figure \ref{fig-naive-diagram}. 

\begin{figure}
    \centering
    \includegraphics[scale=0.35]{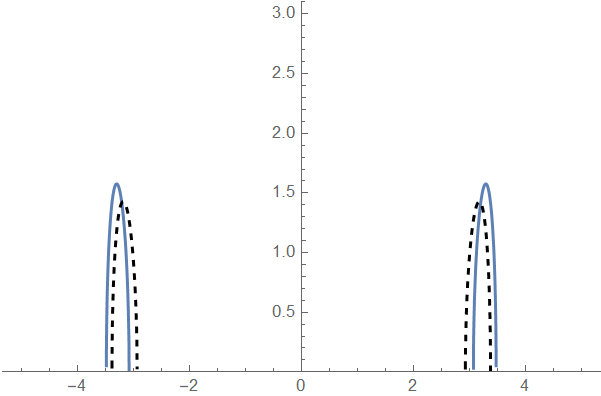}
    \includegraphics[scale=0.35]{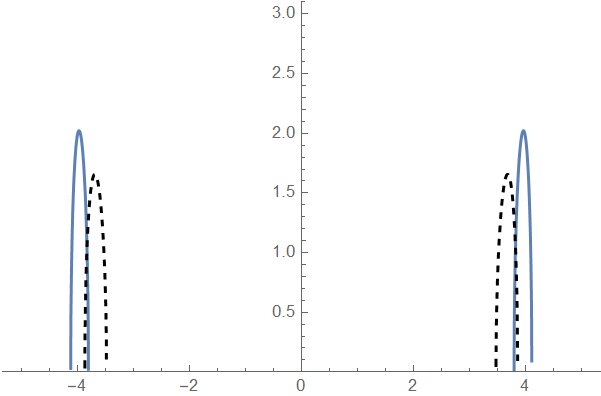}
    \includegraphics[scale=0.35]{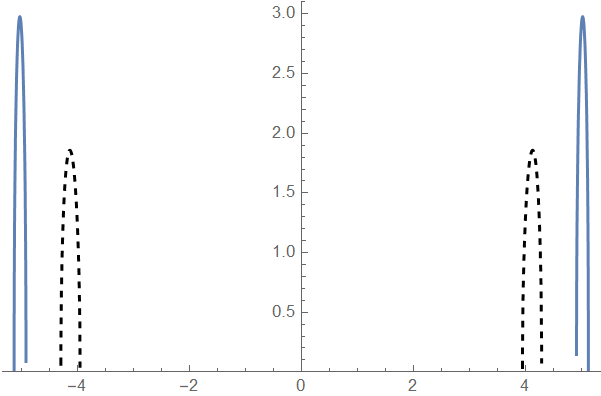}
    \caption{Plots of the eigenvalue distributions of the pure potential matrix model (black and dashed) and the multi-trace matrix model (\ref{5sphere}) for values of ${r=\{-20,-27,-34\}}$ and $g=0.5$. In all the cases we can see that the interaction spreads the eigenvalues further apart. The wider the dashed distribution, the stronger the repulsive interaction spreading the multi-trace distribution.}
    \label{fig-naive-plots}
\end{figure}

One finds out that the two-cut solution ceases to exist in the region of parameter space below the blue line in this figure. This means that if we lower $r$ for a fixed value of $g$, the eigenvalue distribution widens and the repulsive selfinteraction due to the multi-trace terms renders the two-cut solution impossible. This is illustrated in the figure \ref{fig-naive-plots}.

The particular properties of the repulsive self-interaction are however a consequence of the numerical values of the coefficients in (\ref{5sphere}). Namely the coefficient $1/5760$ of the $t_2^4$ term and the coefficient $-1/3456$ of the $(t_4-2t^2_2)^2$ term. Models, where the former coefficient is greater than the absolute value of the latter do not have this problem. But it is still not clear, where and whether at all the asymmetric solution is preferred and such models require further investigation. Since our goal is the analysis of the field theory on the fuzzy sphere, we will not go along this way any further.

\begin{figure}
    \centering
    \includegraphics[width=0.75\textwidth]{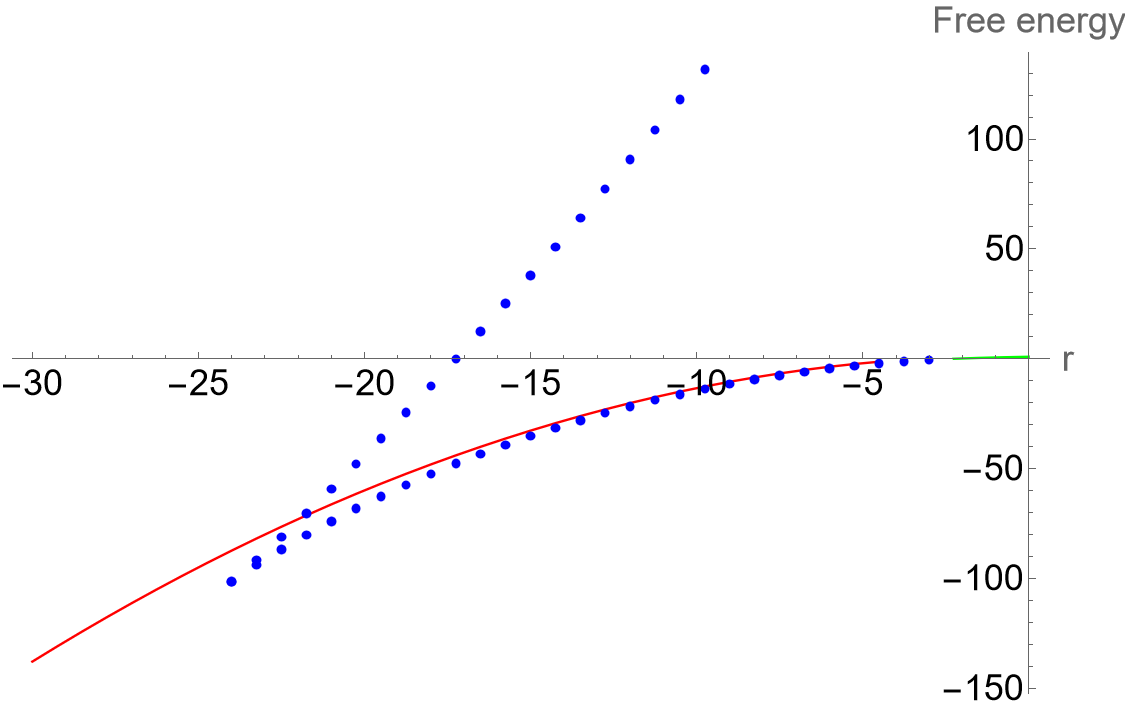}
    \caption{Free energies of the three phases for a fixed value ${g=0.4}$. The color code is the same as in the previous figure: green line is the symmetric one-cut free energy, red line is the asymmetric one-cut free energy and the blue dots are free energies of the two different types of two-cut solution. As we can see, for certain value of $r$ the regime  shifts from the two-cut phase to the asymmetric one-cut phase with a discontinuous jump in the free energy. Note that there are two different two-cut solutions for given values of parameters, only the solution with lower free energy is relevant for our discussion.}
    \label{fig-naive-fixedG}
\end{figure}

\subsubsection*{Phase diagram}

The two-cut solution abruptly ceases to exist along a specific line in the parameter space and the asymmetric one-cut solution takes over, illustrated in the figure \ref{fig-naive-diagram}. In the shaded region in the left image of this figure, both the asymmetric one cut and symmetric two-cut solutions exist, but the two-cut solution has lower free energy and is thus the preferred solution. However at the bottom of the shaded region, along the blue line, the system abruptly shifts from the symmetric two-cut to the asymmetric one-cut regime, with a discontinuous jump in the free energy, as illustrated in the figure \ref{fig-naive-fixedG}. This is a very different behavior than the one expected from a matrix model describing fuzzy field theory, where the phase transition between the two solutions has been observed to be much milder \cite{samo}.

There is a different problem with the phase diagram of the model (\ref{5sphere}) in the vicinity of the origin of the parameter space, shown in the right image of the figure \ref{fig-naive-diagram}. The three lines do not meet at one point and there is no triple point. Moreover, there is a region of the parameter space - the shaded region in the right image - where no solution to the model (\ref{5sphere}) exists. This is again in contradiction to what we expect from the model describing the scalar field theory on the fuzzy sphere. Before wrapping this section up, let us comment on two other features of the phase diagram close to the origin. First is a patch of parameter space, shown by the gray line, where an asymmetric solution to the model exists that is not described by the solution (\ref{asymD}). This solution was however observed by numerical solution of the defining equations (\ref{exp_asym}-\ref{exp_asym0}) and slightly overlaps with the shaded region along the green line, but still leaves a large part of the parameter space without any solution at all. The second feature is that the line (\ref{sec5-full}) is not a phase transition line close to the origin. Two different symmetric one-cut solutions exist in the region between the dashed blue line and the green line\footnote{The reason for this is rather technical and diverts from the main line of our discussion. The second power of the fourth moment of the distribution creates a complicated relation between the true coupling $g$ and the coupling of the effective single trace model $g_e$. As a result, the sheet of the free energy dependence of the effective single trace model gets folded and what used to be a transition line in the original effective model now occurs above other solutions with lower free energy, and thus does not get realized.}. Moreover, symmetric one-cut solution exists for negative values of $r$ all the way to ${g=0}$. This is however also in contradiction to the exact solution of the complete model (\ref{correlation}) in the case of ${g=0}$ in \cite{steinacker05,PNT12}, where it has been shown that the kinetic term only rescales the radius of the symmetric one-cut distribution but does not introduce any new solutions. 

To summarize, the model (\ref{5sphere}) has no chance of describing the numerically observed phase diagram of the scalar field theory on the fuzzy sphere. The asymmetric one-cut solution of the model, which represents the uniform order phase of the field theory, exists in the expected region of the parameter space. However, the two-cut solution does not exist in the expected region of the parameter space and the transition between the symmetric two-cut and the asymmetric one-cut solution is way too abrupt. The free energy discontinuously jumps from one regime to another, opposing to a continuous change seen in the numerical simulation. The model (\ref{5sphere}) does not have a triple point and for some values of the parameters does not have any solution at all.

We need an approximation of the kinetic term effective action (\ref{sec3:effaction}) that is not perturbative in the eigenvalues of the matrix and  would consider the different behavior of this action for large values of the moments, as is the case in the nonperturbative second-moment approximation (\ref{F2}). The two-cut solution would have a chance to exist for arbitrarily large $-r$ and the phase transition to asymmetric one cut solution could be less abrupt.

\section{Extended models for the fuzzy sphere}\label{sec6}

\subsection{General idea}

We now understand quite well where the issues with the model (\ref{5sphere}) are. The large moment behavior of the effective action renders the model unstable. We also understand the origin of the problem. The expression (\ref{5sphere}) is a small moment series of a more complicated expression cut off after the eight order in eigenvalues, the first nontrivial order in which $t_3$ and $t_4-2t_2^2$ appear. But we also understand the possible solution to this issue in the form of the complete $t_2$ dependence of the effective action \cite{MSJT2020,CORFU19a}. The higher terms tame its behavior when viewed as a self-interaction and stabilize the solution.

Our approach will thus be to extend the perturbative, small moment behavior (\ref{5sphere}) with some form of large moment behavior which is much less divergent similarly to the large $t_2$ behavior of (\ref{F2}). Unfortunately no analytical results for the forms of the integral (\ref{angular}) in this regime are available at the moment, so we will try several possible forms by hand and investigate if it is possible to extend the effective action in a way that would lead to a reasonable phase diagram of the model and reasonable location of the triple point.

\subsection{Simple logarithm approximation}\label{sec6.2}


We have seen that the perturbative approximation of the effective action does not overall lead to a phase diagram expected from the fuzzy field theory. We want to include the higher moments into the effective action in such a way that does not spoil the aspects that the second-moment approximation got right but at the same time targets the shortcomings of this nonperturbative approximation. We have seen that in order to do so, we need to go beyond the perturbative expression for the effective action and consider also its large moment behavior. As a first attempt, we get inspiration from the large moment behavior of the second-moment approximation (\ref{F2}) and try the same also for the higher moments.

We thus consider the model given by
\be
S_{eff}=\half \log\lr{\frac{t_2}{1-e^{-t_2}}}+F_3(t_3)+ F_4(t_4-2t_2^2)
\ee
with the functions ${F_3, F_4}$ not needed exactly, as we require only the small and the large parameter expansions in our approach to solution of the model. 

As the first attempt, we follow large expansion of (\ref{F2}) and consider the large expansion of $F_4$ given only by the logarithmic term
\be\label{62_large}
F_4(y_4) = \alpha_0 \log y_4\ ,
\ee
without any further terms. The function $F_3$ is relevant only for the asymmetric solution, where the small expansion of the effective action is used. Therefore we do not need to know its large expansion. 

The small expansion of the effective action will be given by the perturbative series (\ref{5sphere})
\begin{align}
F_3(y_3) = - \frac{1}{432}y_3^2\ ,\ 
F_4(y_4) = -\frac{1}{3456}y_4^2\ .
\end{align}

As in the section \ref{sec5}, we will first compute the symmetric phase transition, followed by the formulae for the symmetric two-cut and asymmetric one-cut solutions, which yield the second phase transition line. The symmetric one-cut solution together with the final transition line require more technical work. Finally, we present the phase diagram and its dependence on the value of the coefficient $\alpha_0$.

\subsubsection*{Symmetric one-cut to symmetric two-cut phase transition}

The symmetric transition line is given by (\ref{sym_trans1},\ref{sym_trans2}). Using the large moment expansions of the functions, which are in this case equivalent to the small $g$ series expansions, we get
\begin{align}
    g_e  = g - \frac{4\alpha_0}{3}g_e\ , \ 
    r = -\frac{8g-4g_e}{\sqrt{g_e}} - \sqrt{g_e}\ ,
\end{align}
which can be combined to  
\be
r = -\frac{15+ 32\alpha_0}{\sqrt{9+12 \alpha_0}}\sqrt{g}\ . \label{sym_transition}
\ee
Note that we have neglected the exponentially suppressed terms in the large expansion of the second moment function (\ref{F2}). W expect this transition to be in the positive $g$, negative $r$ quadrant of the parameter space, which requires ${\alpha_0>-\frac{15}{32}}$. For the values ${\alpha_0<-\frac{15}{32}}$, the repulsion introduced between eigenvalues is strong enough that the one-cut solution ceases to exist even for some positive values of $r$. For ${\alpha_0< - \frac{3}{4}}$,  the repulsion is too strong for the one-cut solution to exist altogether, which can be checked by looking for the solution of the defining equations numerically. 

The further terms of the order $\frac{1}{y_4}$ and higher in the large expansion of $F_4(y_4)$ do not significantly alter the transition line near the origin of the parameter space. Such higher terms introduce only a correction of the order $g^{3/2}$, however, the leading term remains unchanged.  We discuss this more thoroughly in Section \ref{sec6.3}.

\subsubsection*{Symmetric two-cut to asymmetric one-cut phase transition}

The transition is obtained from the condition
\begin{equation}
\mathcal{F}_{\textrm{as1c}} - \mathcal{F}_{s2c} = 0\ . \label{con_as1c-s2c}
\end{equation}
We obtain the corresponding free energies by solving the equations for the symmetric two-cut (\ref{exp_2cut},\ref{exp_2cut0})  and the asymmetric phase (\ref{exp_asym}-\ref{exp_asym0}) as a series expansions in $1/r$.

For the two-cut phase, the large parameter expansions of the moment functions are relevant, and we get
\begin{align}
D & = - \frac{r}{4g} + \frac{1+4\alpha_0}{r}+ \frac{4 (1 + 12 \alpha_0 + 16 \alpha_0^2) g}{r^3}+ \dots\ , \\
\delta & = \frac{1}{\sqrt{g}}+ \frac{8\alpha_0\sqrt{g}}{r^2} + \frac{32 (3 \alpha_0 + 11 \alpha_0^2)g^{3/2}}{r^4} + \ldots\ , \\
\mathcal{F}_{s2c} & =- \frac{r^2}{16g} + \bigg(\frac{3}{8} +\frac{1+4\alpha_0}{2} \log (-r)  - \frac{1+8\alpha_0}{4} \log (4g) \bigg) - \frac{(1 + 12 \alpha_0 + 16 \alpha_0^2) g}{r^2} + \ldots\ .\label{F_s2c}
\end{align}

For the asymmetric phase, the small parameter expansions of the functions are relevant, and we obtain the same results (\ref{asymD}-\ref{F_as1c}) as in the section \ref{sec5}, since we are working with the same form of the action. We solve the condition (\ref{con_as1c-s2c}) perturbatively as well, leading to the following expression for the phase transition in the leading order
\begin{equation}
g= \frac{1}{4}e^{\frac{-(3+ \log 16)}{2(1+8\alpha_0)}}(-r)^{\frac{8\alpha_0}{1+8\alpha_0}}\ . \label{as1c-s2c_lowest}
\end{equation}
This expression is valid only for ${\alpha_0> - 1/8}$. Else, the exponent of $|r|$ is greater than $1$ and the higher orders of the perturbative solution of (\ref{con_as1c-s2c}) are not progressively smaller. However, the numerical solution of the corresponding equations suggests that below this value, not only our perturbative approach fails, but the phase transition ceases to exist altogether, as the asymmetric phase has always higher free energy than the two-cut phase. This can be interpreted that for ${\alpha_0< - 1/8}$, the repulsion introduced between eigenvalues becomes strong enough for the two-cut solution to be always preferred over the asymmetric one-cut solution. 

Note that for ${\alpha_0=0}$, we get $g$ equal to a constant, which is the large $|r|$ asymptotic behavior of the transition line in the pure second-moment model discussed in the Section \ref{sec3.2}. For the negative $\alpha_0$, the asymmetric phase region shrinks with the increasing $|r|$, as seen in the first diagram in the figure \ref{fig-c0-1/24}. For the positive values of $\alpha_0$, the contrary is true. For the transition to behave linearly, as is suggested by the numerical simulation, one would need to take ${\alpha_0 \rightarrow \infty}$.  However, the exponent in (\ref{as1c-s2c_lowest}) reaches quite soon values that would be very difficult to distinguish from the linear behavior in numerical results.  

The following general expression for the higher orders of the phase transition (\ref{as1c-s2c_lowest}) can be obtained in case of ${\alpha_0>0}$: 
\begin{align}
g \, &= \,  \frac{1}{4}e^{\frac{-3- \log 16}{2+16\alpha_0}}(-r)^{\frac{8\alpha_0}{1+8\alpha_0}}-\frac{1}{8(8\alpha_0+1)}e^{\frac{-3- \log 16}{2+16\alpha_0}}(-r)^{\frac{8\alpha_0}{1+8\alpha_0}-1}- \nonumber \\
&+  \frac{3-24\alpha_0-32\alpha_0^2}{8(8\alpha_0+1)}\,e^{\frac{-3- \log 16}{1+8\alpha_0}}(-r)^{2(\frac{8\alpha_0}{1+8\alpha_0}-1)}+\ldots\ . \label{as1c-s2c}
\end{align}
As the exponent in the leading order is not necessarily an integer number, the higher orders contain the terms with the exponent decreasing by one as well as all the multiples of such exponents. We Pad\'e approximate the transition $(\ref{as1c-s2c})$ to treat the divergent behavior of the series at the origin, analogous to what was done for the second-moment model in \cite{MSJT2020}. However, in the case of non-zero $\alpha_0$, we deal with the various rational exponents in the series. Therefore, we are generally able to obtain the Pad\'e approximation only of a very low order.

\subsubsection*{Symmetric one-cut to asymmetric one-cut phase transition}

This transition line is acquired from the condition
\begin{equation}
\mathcal{F}_{as1c} - \mathcal{F}_{s1c} = 0\ . \label{con_as1c-s1c}
\end{equation} 
As the symmetric one-cut phase does not exist in the region of the large $|r|$ for some constant $g$, one needs to slightly modify the perturbative approach to obtain this solution and the phase transition \cite{MSJT2020}.

We expand the equations for the symmetric one-cut phase (\ref{exp_sym},\ref{exp_sym0}) around the symmetric phase transition (\ref{sym_transition}) using the large expansion of the effective action ${S_{eff}}$ and obtain the following solution
\begin{align}
\delta =& \frac{2\sqrt{3+4\alpha_0}}{\sqrt{3g}} + \frac{27 - 24 \alpha_0 - 80 \alpha_0^2}{(-81 + 132 \alpha_0 + 128 \alpha_0^2) g} \bigg( r+ \frac{(15 + 32 \alpha_0) \sqrt{g}}{\sqrt{9+12\alpha_0}}\bigg) + \ldots\ , \label{s1c_delta} \\
g_e =& \frac{3g}{3+4\alpha_0}+ \frac{90\alpha_0 \sqrt{3g}}{\sqrt{3+4\alpha_0}(-81 + 132 \alpha_0 + 128 \alpha_0^2)} \bigg( r+ \frac{(15 + 32 \alpha_0) \sqrt{g}}{\sqrt{9+12\alpha_0}}\bigg) + \ldots\ , \\
F_{s1c} =& \alpha_0 \log \bigg ( \frac{3+4\alpha_0}{g} \bigg) + \frac{\log (16)(3-8\alpha_0)-27-88\alpha_0}{24}+\nonumber\\&+ \frac{\sqrt{3+4\alpha_0}}{2\sqrt{3g}}\bigg( r+ \frac{(15 + 32 \alpha_0) \sqrt{g}}{\sqrt{9+12\alpha_0}}\bigg) + \ldots . \label{s1c_free}
\end{align} 
Note that the symmetric one-cut to asymmetric one-cut phase transition ends in the triple point of the theory. It, therefore, lies in the region ${g<g_c}$, ${|r|<|r_c|}$. The critical point in the second-moment model (i.e. ${\alpha_0 = 0}$) lies very near the origin of the parameter space and this fact justifies the use of the large series expansions of the moment functions, as for small values of $g$ the moments are large. This also holds for the small values of $|\alpha_0|$. However, as we will discuss in more detail later, for the larger positive values of $\alpha_0$ the triple point moves significantly further from the origin, making our perturbative approach no longer valid.  

To obtain the symmetric one-cut to asymmetric one-cut phase transition from the condition (\ref{con_as1c-s1c}) we first need to adjust the expression for $F_{as1c}$ (\ref{F_as1c}) as it is the expansion around different region of the parameter space. We Pad\'e approximate the $O(1/r)$ part of the  $F_{as1c}$ series expansion and then re-expand the obtained expression around the symmetric transition (\ref{sym_transition}) \cite{MSJT2020}.

The phase transition is then obtained using final few technical tricks. First, we replace
\begin{align*}
g = \frac{g_c}{(1+x)^2}\ ,
\end{align*} 
with the triple point value $g_c$ calculated numerically as the intersection of Pad\'e approximated (\ref{con_as1c-s1c}) with (\ref{sym_transition}). We then solve the condition (\ref{con_as1c-s1c}) order by order, obtaining the transition line in the form
\begin{align}
r(x) = r_0 + r_1 x + r_2 x^2 + r_3 x^3 + \ldots\ . \label{as1c-s1c_trans}
\end{align} 
Note that $r_i$ also depends on the order of the calculation. The coefficients change with the order of calculation and this change needs to be less and less significant with the increasing order for the series to converge. 

We then Pad\'e approximate this expansion in $x$ taking into account that we expect the transition goes through the origin of the parameter space.

In this manner, we obtain a reasonable phase transition for the small values of $\alpha_0$. 
\subsubsection*{Phase diagram and triple point}

\begin{figure}
    \centering
    \includegraphics[scale=0.5]{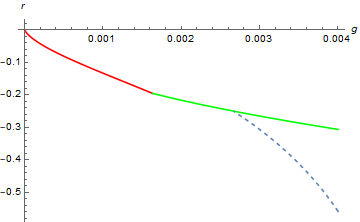}
    \includegraphics[scale=0.6]{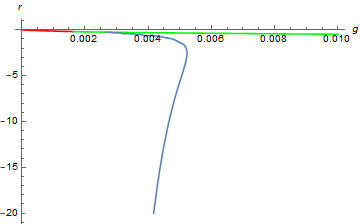}
    \caption{The phase diagram in case of $\alpha_0= -1/48$. The blue line corresponds to the asymmetric one-cut to the symmetric two-cut transition, the red line to the asymmetric one-cut to the symmetric one-cut transition and the green line to the phase transition between the symmetric phases. The asymmetric one-cut to the symmetric two-cut phase transition on the left image is plotted in the dashed style as this transition is less accurate near the origin of the parameter phase due to the order of the Pad\'{e} approximation. The triple point was obtained at $g_c =  0.0016$. }
    \label{fig-c0-1/24}
\end{figure}

\begin{figure}
    \centering
    \includegraphics[scale=0.5]{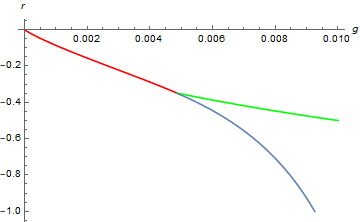}
    \includegraphics[scale=0.6]{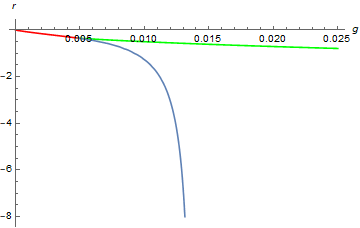}
    \caption{The phase diagram in case of $\alpha_0= 0$. The blue line corresponds to the asymmetric one-cut to the symmetric two-cut transition, the red line to the asymmetric one-cut to the symmetric one-cut transition and the green line to the phase transition between the symmetric phases. The triple point was obtained at $g_c =  0.0048$.}
    \label{fig-c0-0}
\end{figure}

\begin{figure}
    \centering
    \includegraphics[scale=0.5]{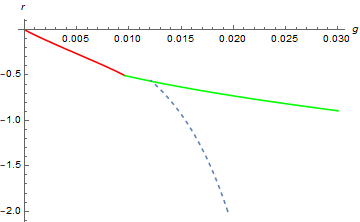}
    \includegraphics[scale=0.6]{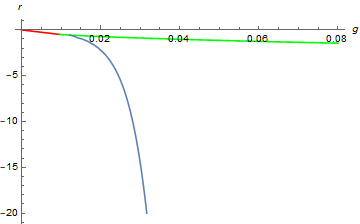}
    \caption{The phase diagram in case of $\alpha_0= 1/48$. The blue line corresponds to the asymmetric one-cut to the symmetric two-cut transition, the red line to the asymmetric one-cut to the symmetric one-cut transition and the green line to the phase transition between the symmetric phases. The asymmetric one-cut to the symmetric two-cut phase transition on the left image is plotted in the dashed style as this transition is less accurate near the origin of the parameter phase due to the order of the Pad\'{e} approximation. The triple point was obtained at $g_c =  0.0096$.}
    \label{fig-c01/24}
\end{figure}

\begin{figure}
    \centering
    \includegraphics[scale=0.5]{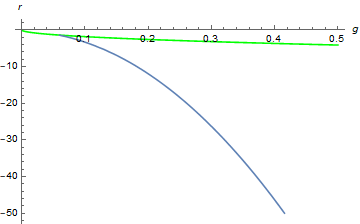}
    \includegraphics[scale=0.5]{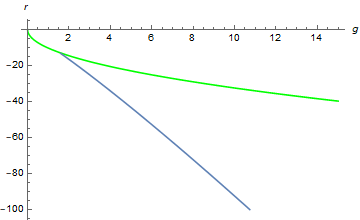}
    \caption{The phase diagrams for the value $\alpha_0= 1/8$ (left) and $\alpha_0= 1$ (right). The blue line corresponds to the asymmetric one-cut to the symmetric two-cut transition and the green line to the transition between the symmetric phases. The triple points were determined at $g_c=0.054$ and $g_c=1.43$ respectively.}
    \label{fig-c01}
\end{figure}

The phase diagrams for the different values of $\alpha_0$ are pictured in the Figures \ref{fig-c0-1/24}-\ref{fig-c01}. The asymmetric one-cut to the symmetric one-cut transition line was calculated up to the eighth order in $\sqrt{g}$, and then Pad\'e approximated. In the case of the asymmetric one-cut to the symmetric one-cut phase transition, we were generally able to obtain only the second-order Pad\'e approximation. Therefore, the behavior of this transition near the origin is less accurate. 

The triple point was obtained numerically as the intersection of the symmetric phase transition (\ref{sym_transition}) with the asymmetric one-cut to the symmetric one-cut phase transition (\ref{as1c-s1c_trans}). For the larger positive values of $\alpha_0$ when the asymmetric one-cut to the symmetric one-cut transition line was not obtainable, the triple point was determined as the intersection of the symmetric transition and the Pad\'e approximated asymmetric one-cut to the symmetric two-cut transition (\ref{as1c-s2c}).

We can observe the general trend of moving the triple point further from the origin with increasing $\alpha_0$. The explicit calculation of the intersection between the symmetric phase transition (\ref{sym_transition}) and the lowest order of the asymmetric one-cut to symmetric two-cut transition (\ref{as1c-s2c_lowest}) gives
\begin{align}
 g_c & = \frac{1}{16} \bigg[ \frac{1}{e^{3/2}} \bigg( \frac{1}{81}\bigg)^{\alpha_0} \bigg(\frac{15+32\alpha_0}{\sqrt{3+4\alpha_0}}\bigg)^{8\alpha_0} \bigg]^{\frac{1}{1+4\alpha_0}}\ , \\
 r_c &  = - \frac{15+32\alpha_0}{4\sqrt{9+12\alpha_0}}  \bigg[ \frac{1}{e^{3/2}} \bigg( \frac{1}{81}\bigg)^{\alpha_0} \bigg(\frac{15+32\alpha_0}{\sqrt{3+4\alpha_0}}\bigg)^{8\alpha_0} \bigg]^{\frac{1}{2+8\alpha_0}}\ .
\end{align}
Although the higher orders adjust these values quite significantly, the general trend of increasing $g_c$, $|r_c|$ with the raising of $\alpha_0$ remains the same. Note that this is in contrast with the numerical simulations, with the value of $r_c$ significantly closer to the origin than in the second-moment approximation. This would indicate that if we had only (\ref{62_large}), we would need a negative value of $\alpha_0$ to bring the triple point closer to the numerically observed value. 

The large moment behavior (\ref{62_large}) of the effective action did lead to a phase diagram similar to the one expected for the fuzzy field theory for positive $\alpha_0$. It solved the most important problem of the previous second-moment models and leads to a very reasonable asymmetric one-cut to symmetric two-cut phase transition line. But such $\alpha_0$ takes the triple point even further away from the numerically observed location. We thus need to consider a more complicated behavior of the kinetic term effective action.

\subsection{Other possible extensions} \label{sec6.3}

To conclude this report, we would like to investigate the consequences of some different and more elaborate extensions.

\subsubsection*{Beyond pure logarithm}

We consider the higher order terms in the large expansion of the function $F_4(y_4)$
\begin{align}
F_4(y_4) = \alpha_0\log(y_4) + \alpha_1+ \frac{\alpha_2}{y_4}+ \frac{\alpha_3}{y_4^2}+ \ldots\ . \label{f4_beyond}
\end{align} 

Such a model can be treated in the same way as we described in the previous section with little modification. The asymmetric one-cut to the symmetric two-cut phase transition can be obtained following the same procedure stated in the previous section. We obtain the leading order of the form
\begin{align}
g = e^{-\frac{3+\log 16-8\alpha_1}{2+16\alpha_0}}(-r)^{\frac{8\alpha_0}{1+8\alpha_0}}\ .
\end{align}
We can see that besides the parameter $\alpha_0$, only the constant term $\alpha_1$ appears in the leading order. However, only $\alpha_0$ affects the exponent of $r$. Therefore, the higher-order terms in (\ref{f4_beyond}) do not alter the character of the phase transition for large values of $|r|$, as this depends only on the exponent value. The constant term $\alpha_1$ is added to the free energy of the two-cut solution. Therefore, it extends or shrinks the region of this phase according to its sign. 

In the case of the symmetric phase transition, we are no longer able to obtain this line in the large regime exactly. Instead, we get the perturbative expression in $\sqrt{g}$: 
\begin{align*}
r =& -\frac{15+32\alpha_0}{\sqrt{9+12\alpha_0}}\sqrt{g}- \frac{8\alpha_2(33 + 32 \alpha_0)}{\sqrt{3}(3 + 4 \alpha_0)^{5/2}}g^{3/2}+ \nonumber \\
&+\frac{32\big((183 + 160 \alpha_0) \alpha_2^2 + 2 (99 + 228 \alpha_0 + 128 \alpha_0^2) \alpha_3 \big)}{\sqrt{3}(3 + 4 \alpha_0)^{9/2}}g^{5/2} + \ldots\ .
\end{align*}
We can see that the constant term $\alpha_1$ does not affect this transition at all, as we use only the derivative of $F_4$ in the calculation. This is understandable as we are adding the same constant to both free energies. 

To obtain the third transition line, asymmetric one-cut to symmetric one-cut phase transition, we need to expand the corresponding equations around the symmetric transition. However, since we now know the symmetric transition only perturbatively, the coefficients in the expansions equivalent to (\ref{s1c_delta}-\ref{s1c_free}) will be also known only perturbatively in the small $g$. 

As we mentioned, the constant terms $\alpha_1$ adds this value to the free energy of the symmetric solutions and can therefore significantly alter the location of the triple point. However, the triple point still must be located on the symmetric transition, which is unaffected by $\alpha_1$.

Thus, to set the triple point to any desired location, e.g. to the location obtained in the most recent numerical simulations, we first need the symmetric transition to pass through this point. However, we can alter the symmetric line with the higher order terms only to a limited degree. If we were to obtain the exact solution of the equation (\ref{sym_trans1}), which is in general possible only up to the $\alpha_4$ order, we would see that this equation does not always have a solution for the larger values of $g$. For which particular value of $g$ the solution ceases to exist depends on the specific choice of $\alpha_i$.

Therefore, we cannot force the symmetric transition to pass through an arbitrary point. For example, if we look at $g_c= 0.0048$ and set $\alpha_0 \geq 0$, we cannot bend the transition line enough to pass through as small values of $|r|$ as obtained in the numerical simulation because the solution ceases to exist.

\subsubsection*{Logarithm-less modifications}

To conclude this report, we discuss the large $y_4$ behavior of the functions $F_4$ different from $\log y_4$. The question is what kinds of behaviors lead to a reasonable phase transition between the two-cut and the asymmetric one cut solutions. From the free energies of the two solutions (\ref{F_s2c}) and (\ref{F_as1c}) we see that the leading change to the two-cut free energy needs to be at $\log(-r)$ or $r^2$ terms. Anything else leads to ill behaved models with at best very abrupt phase transitions, as the one discussed in Section \ref{sec5}, or no phase transitions at all.

This can be explicitly seen in the model (\ref{f4_beyond}) with $\alpha_0=0$. Without the freedom in the $\alpha_0$ parameter, we cannot alter the large $|r|$ behavior of the asymmetric one-cut to the symmetric two-cut phase transition. Therefore, we are unable to address the discrepancy between the numerical results and the second-moment models (\ref{F2}) at all within the scope of this model.

We can, however, alter the location of the triple point.  The value $g_c$ of the second-moment model corresponds with the value obtained in the numerical simulations. The $|r_c|$ value is approximately ten times larger than the value from the numerical simulations. By suitable choice of the higher-order parameters, we can alter the $r_c$ value, moving it closer to the origin. However, as was also the case in the previous section, we are not able to set it small enough to agree with the numerical data. The smallest value obtainable by the higher-order terms in the large expansion is still about $2.5$ times larger than the value suggested numerically.

Finally, what about the models where $F_4$ grows faster than logarithm? The calculations in this case are very similar to what has been done in Section \ref{sec5}, with an important difference in the large $t_2$ behavior of the second-moment part of the effective action. And it can be shown that for any effective action of the for\footnote{The absolute value is used, since $t_4-2t_2^2$ is negative in the two-cut regime.} $F_4=\beta |t_4-2t_2^2|^n$, for $n>1$ there is no large $-r$ solution to equations (\ref{exp_2cut}-\ref{exp_2cut0}). The self interaction is either too repulsive or two attractive to support a stable two-cut solution. For $n=1$ the solution does exist, but its free energy is always larger or smaller, depending on sign of $\beta$, and we are thus left with no phase transition.

For the case ${0<n<1}$, we do obtain a stable two-cut solution, and for attractive force $\beta>0$ even a phase transition given by, in the leading order,
\begin{align}
    g=2^{\frac{3}{2 n}-2} \left(\frac{\beta n}{W\big(8 (48)^n\beta  n (-r)^{2 n}\big)}\right)^{\frac{1}{2 n}}(-r) \ ,\label{sec6-leading}
\end{align}
where $W(z)$ is the product logarithm or the Lambert $W$ function. This behavior can, for particular values of $n$ and $\beta$, be similar enough to linear behavior and is thus not inconsistent with the available numerical data. For the symmetric phase transition, the large parameter expansion of the effective action is relevant close to the origin of the parameter space and as we will shortly see, this behavior will not be needed.

Due to the form of the leading order contribution (\ref{sec6-leading}) it is not possible to compute further $1/r$ contributions to the asymmetric phase transition line even for the simplest case of $n=1/2$. It is however possible to investigate the solution of the equations from Section \ref{sec4.1}, including the free energies and the transition lines between them, numerically. For this square root effective action we find out that the transition line is very well described by the leading order contribution and we can locate the triple point as an intersection of (\ref{sec6-leading}) and (\ref{sec5-full}). Even for a small value ${\beta=10^{-3}}$ we obtain a location, that is significantly further away from the value observed in the numerical simulations than the result for the second-moment model.

For different values of $n$ the numerical computation shows that the triple point moves towards the origin as $n$ approaches zero. However to obtain a value that is reasonably close to the value observed in numerical simulations, we would need an unreasonably small $n$. We thus conclude that effective actions of other than logarithmic behaviour for large ${t_4-2t_2^2}$ introduce an interaction that is too strong and does not lead to a phase diagram consistent with numerical simulations.

\section{Conclusions}\label{sec_conclusions}

We have analyzed the fourth-moment fuzzy-field-theory-like matrix models and their phase structure. We have shown that combining a perturbative term proportional to the symmetrized third and fourth moments of the eigenvalue distribution of the matrix with a logarithmic fourth-moment term in the large moment regime, which we added by hand, leads to a phase diagram consistent with the most recent numerical simulations. It fixes the most important shortcoming of the previous second-moment models and can lead to a roughly straight transition line between the asymmetric one-cut and the symmetric two-cut phases.

Our analysis confirms that looking at the small moments expansion of the kinetic term effective action is not sufficient and the large moment behavior is crucial for the analysis of the phase structure, most importantly to explain the region of stability of the asymmetric phase and existence of the triple point.

We were however unable to match the location of the triple point of the studied model to the results of the numerical simulations. Even after the addition of subleading terms, there is no choice of the coefficients - the free parameters in the studied models - that would lead to a location of the triple point consistent with the numerically observed value. This means that fourth-moment models are not sufficient to explain quantitatively the phase structure of the fuzzy field theories completely and to do so, one needs to go even further.

It would be very interesting to calculate the large moment behavior of the kinetic term effective action directly, to verify the leading logarithmic behavior, and to see what the value of the coefficients is. This calculation could also shed some more light on the properties and structure of the dependence of the effective action on the eigenvalues of the matrix. Perhaps there is a better way to express this dependence in the large eigenvalue regime than the symmetrized moments expansion used in the small moment regime.

Finally, there are several results for the fuzzy field theories that go beyond the phase structure of the model. Most notably the properties of the correlation functions \cite{correlationFunctions0,correlationFunctions1,correlationFunctions2} and the entanglement entropy \cite{entanglement1,entanglement2,entanglement3,entanglement4} have been investigated. A very natural next step would be to analyze these also using the above methods of the multi-trace matrix models.

\acknowledgments
We would like to thank Samuel Kov\'a\v cik and Peter Pre\v snajder for many helpful discussions.

This work was supported by the \emph{Alumni FMFI} foundation as a part of the \emph{N\'{a}vrat teoretikov} project and by VEGA 1/0703/20 grant \emph{Quantum structure of spacetime}.


\appendix
\section{Technical details for the perturbative calculations}\label{app_direct}

This appendix contains the technical details involved in the derivation of the conditions on the distributions given in the section \ref{sec4.1}.

We deal with the matrix model
\be\label{app_action}
S=\half r c_2+g c_4+F\slr{c_1,t_2,t_3,t_4-2t_2^2}\ , 
\ee
where the moments of the eigenvalue distributions are
\begin{align}
c_n=\int dx\,x^n\rho(x)=\frac{1}{N}\trl{M^n}
\end{align}
and the symmetrized moments are
\begin{align}
t_n\,=&\, \frac{1}{N}\textrm{Tr}\,\lr{M-\mathbb{1}\frac{1}{N}\textrm{Tr}\,M}^n\ ,\\
t_2 = c_2-c_1^2\ ,\ 
t_3 =\,& c_3-3c_1c_2+2c_1^3\ ,\ 
t_4 = c_4-4c_1c_3+6c_2c_1^2-3c_1^4\ .
\end{align}
As before we denote
\begin{align}
f_1=\pd{F}{y_1}\ ,\ f_2=2\pd{F}{y_2}\ ,\ f_3=\pd{F}{y_3}\ ,\ f_4=\pd{F}{y_4}\ .
\end{align}
The variation of the action (\ref{app_action}) is
\begin{align*}
\frac{\partial S}{\partial x_i}\, =&\, rx_i+4gx_i^3+f_1 + f_2 (x_i-c_1)+f_3(3x_i^2-6c_1 x_i-3c_2+6c_1^2)- \\&
+f_4\Big(\Big( 4 x_i^3-4(3x_i^2 c_1+c_3)+6(2 x_i c_1^2+2 c_2 c_1)-12 c_1^3\Big)-4 (c_2-c_1^2)2(x_1-c_1)\bigg)
\end{align*}
We write this as a variation of an action
\footnote{A different approach would be to shift the eigenvalues by a constant $x\to x-x_0$ to cancel the cubic term $x^3$ in the effective model action. One would not need to compute a more complicated formula for the effective action of an asymmetric quartic matrix model with a cubic term, but the moments of the effective model would be different from the moments of the true model. We chose to prefer this simplicity and did the general calculation of the free energy.}
\be\label{app_effmodel}
\Seff=\aeff c_1+\half \reff c_2+\beff c_3+\geff c_4 
\ee
and we read of
\begin{align}
\aeff & =  f_1-f_2 c_1-3f_3(c_2-2c_1^2)-4f_4(c_3-5c_1c_2+5c_1^3)\ ,\label{app_first}\\
\reff & = r+ f_2-6f_3 c_1-4f_4(2c_2-5c_1^2)\ ,\\
\beff & = f_3- 4 f_4 c_1\ , \\
\geff & =  g+f_4\label{app_last}\ .
\end{align}

The plan of attack is now the following. Model (\ref{app_effmodel}) can be solved using the standard methods \cite{matrixmodels,cm} and the distribution, and thus also the moments $c_n$, obtained in terms of ${\aeff,\reff,\geff}$. These however depend on the moments $c_n$ through the conditions (\ref{app_first}-\ref{app_last}). These then supplement the conditions on the parameters of the distribution in the solution of the model (\ref{app_effmodel}) in defining the solution to the model (\ref{app_action}).

We will give the explicit formulae for the cases of symmetric one-cut and two-cut solutions and an asymmetric one-cut solution, which were shown and used in Section \ref{sec4}.

\subsubsection*{One-cut solutions}
A one-cut solution, symmetric or asymmetric, is supported on the interval ${(D-\sqrt{\delta},D+\sqrt{\delta})}$ determined by the conditions
\begin{align*}
0 &= \frac{1}{2}\aeff+\frac{3 \beff D^2}{2}+\frac{3 \beff\delta}{4}+2D^3\geff+3D\delta \geff+\frac{1}{2}D\reff\ , \\
1&=\frac{3 \beff D \delta}{2}+ 3D^2\delta g_e+\frac{3}{4}\delta^2 g_e+\frac{1}{4}\delta r_e\ .
\end{align*}
These can be solved to give $\aeff$ and $\reff$ in terms of the rest of the parameters ${D,\delta,\beff,\geff}$
\begin{align}
\aeff\,=&\,3 D^2 \beff-\frac{3}{2}\delta \beff-\frac{4 D}{\delta} + 8 D^3 \geff - 3 D \delta \geff\ ,\label{app_areff}\\
\reff\,=&\,-6 D \beff+\frac{4}{\delta} - 12 D^2 \geff - 3 \delta \geff\ .\label{app_areff1}
\end{align}
This turns the logic of the equations around but proves to be very useful later. The expressions for the moments simplify to
\begin{align}
c_1\,=&\, \frac{3}{16}\delta^2 \beff +D + \frac{3}{4} D \delta^2 \geff\ ,\label{app_cneff0}\\
c_2\,=&\, 
 \frac{3}{8}D \delta^2\beff +D^2 + \frac{1}{4}\delta + \frac{3}{2} D^2 \delta^2 \geff + \frac{1}{16}\delta^3 \geff\ ,\\
c_3\,=&\,
 \frac{9}{16} \beff D^2 \delta^2+\frac{3}{32} \delta^3 \beff +D^3 + \frac{3}{4} D \delta + \frac{9}{4} D^3 \delta^2 \geff + 
  \frac{9}{16} D \delta^3 \geff\ ,\\
c_4\,=& \,
 \frac{3}{4}\beff D^3\delta^2+\frac{3}{8}\beff D \delta^3+D^4 + \frac{3}{2} D^2 \delta + \frac{1}{8}\delta^2 +\nonumber\\&+ 3 D^4 \delta^2 \geff + 
  \frac{15}{8} D^2 \delta^3 \geff +\frac{3}{64} \delta^4 \geff\ .\label{app_cneff1}
\end{align}

So at the end of the day, we are left with just four equations for  four unknowns - $D,\delta,\beff,\geff$ or $D,\delta,c_1,\geff$. These are (\ref{app_first}-\ref{app_last}), with the use of (\ref{app_areff},\ref{app_areff1}) and the above expressions for $c_n$'s. Explicit form of these equations is given in the main text in the section \ref{sec4.1}. As mentioned before, ${f_1,f_2,f_3,f_4}$ are functions of ${c_1,c_2,c_3,c_4}$ and before proceeding to solution of these equations, one needs to use (\ref{app_cneff0}-\ref{app_cneff1}). In the case of the symmetric solution these simplify considerably to (\ref{exp_sym},\ref{exp_sym0}).

\subsubsection*{Two-cut solution}
We will consider only the case of a symmetric two-cut solution, since we will not encounter such solutions and there is very little one can do analytically for the asymmetric two-cut case. The distribution is supported on the intervals ${(-\sqrt{D+\delta},-\sqrt{D-\delta})\cup(\sqrt{D-\delta},\sqrt{D+\delta})}$ determined by
\begin{align}\label{app_2cut}
r_e = -4Dg_e\ ,\  \delta= \frac{1}{\sqrt{g_e}}\ ,
\end{align}
with moments
\begin{align}
c_2  = D\ ,\ 
c_4  = D^2+ \frac{1}{4g_e}\ .
\end{align}
using this in the expressions (\ref{app_first}-\ref{app_last}) yields the final conditions
\begin{align}
\frac{1}{\delta^2}\,=\,&g+f_4\ ,\\
\frac{4D}{\delta^2}\,=\,&r+8Dg+f_2\ .
\end{align}
If $f_2$ does not involve $t_4$, i.e. if $y_2$ and $y_4$ do not couple in $F$, the second equation can be solved for $\delta$ and one needs to solve only one equation (for $D$) and $\delta$ is given explicitly.

\subsubsection*{Free energies}
The free energy of any distribution $\rho(x)$ is
\begin{align}\label{app_F}
\F\,=\,&S[\rho(x)]-\textrm{Vandermonde term}=\nonumber\\=\,&F\slr{c_1,t_2,t_3,t_4-2t_2^2}+\half r c_2+g c_4-\int dx\,dy\,\rho(x)\rho(y)\log|x-y|\ .
\end{align}
The double integral could be problematic, but fortunately it has been done for us long ago in the computation of the free energy of the effective model (\ref{app_effmodel})
\be
\F_e=\aeff c_1+\half \reff c_2+\beff c_3+\geff c_4-\int dx\,dy\,\rho(x)\rho(y)\log|x-y|\ .
\ee
We obtain
\be 
\F=\F_e+F\slr{c_1,t_2,t_3,t_4-2t_2^2}+\half \lr{ r-\reff} c_2+\lr{g-\geff} c_4-\aeff c_1-\beff c_3\ .
\ee
With expressions for the effective free energies and the moments, we get the formulae for the free energies (\ref{free_1cut}),(\ref{free_2cut}) and (\ref{free_asym}).

\end{document}